\documentstyle[aps,pre,multicol,epsf]{revtex}
\def\gs{\gtrsim}
\def\ls{\lesssim}
\def\be{\begin{equation}}
\def\en{\end{equation}}                  
\newcommand{\bi}[1]{\mbox{\boldmath$#1$}}

\begin{document}
\draft
\bibliographystyle{prsty}
\title{Apparent finite-size effects in the dynamics of supercooled liquids}
\author{Kang Kim$^{1}$ and Ryoichi Yamamoto$^{2}$}
\address{
$^{1}$Department of Applied Mathematics and Physics, 
Graduate School of Informatics, \\
Kyoto University, Kyoto 606-8501, Japan\\
$^{2}$Department of Physics, Kyoto University, Kyoto 606-8502, Japan
}
\date{\today}
\maketitle

\begin{abstract}
Molecular dynamics simulations are performed for a supercooled 
simple liquid with changing the system size from $N=108$ to $10^4$ to 
examine possible finite-size effects.
Although almost no systematic deviation is detected
in the static pair correlation functions, 
it is demonstrated that the structural $\alpha$ relaxation in 
a small system becomes considerably slower than that in larger 
systems for temperatures below $T_c$ at which the size of the 
cooperative particle motions becomes comparable to the unit 
cell length of the small system.
The discrepancy increases with decreasing temperature.
\end{abstract}
\pacs{PACS numbers: 64.70.Pf, 66.10.Cb, 61.43.Fs}

\begin{multicols}{2}

%\section{introduction}

As liquids are cooled toward the glass transition temperature $T_g$,
a drastic slowing-down occurs in dynamical properties, 
such as the structural relaxation time, the diffusion constant, and
the viscosity \cite{Jackle,Tohwa}, while only small changes are detected
in static properties.
The goal of theoretical investigations on the glass transition is to 
understand the universal mechanism which gives rise to the drastic 
slowing-down.
To this end, a great number of molecular dynamics (MD) simulations 
has been carried out for supercooled liquids \cite{Bernu}.
Several large scale simulations have been performed very recently and 
reviled that the dynamics in supercooled liquids are spatially 
{\it heterogeneous}
\cite{Yamamoto1,Yamamoto3,Muranaka,Perera,Kob,Donati};
rearrangements of particle configurations in glassy states occur 
cooperatively involving many molecules.
We have examined bond breakage processes among adjacent particle pairs 
and found that the spatial distribution of broken bonds in an
appropriate time interval 
($\sim \tau_{\alpha}\simeq0.1\tau_b$ where $\tau_{\alpha}$ is the 
structural $\alpha$ relaxation time and $\tau_b$ is the average bond 
life time) 
is very analogous to the critical fluctuation in Ising spin systems.
The structure factor is excellently fitted to the Ornstein-Zernike 
form \cite{Yamamoto1}, and the correlation length $\xi$ thus 
obtained grows rapidly with decreasing temperature.
Furthermore, we demonstrated that $\xi$ is related to $\tau_\alpha$ 
trough the dynamical scaling law, 
$\tau_{\alpha} \sim \xi^z$ with $z\simeq4$ in 2D and $z\simeq2$ in 3D. 
The heterogeneity structure in our bond breakage is essentially the same 
as that in local diffusivity \cite{Yamamoto3}, which leads to a 
systematic violation of the Stokes-Einstein low in supercooled states.

To investigate long-time behavior of glassy materials by MD simulation, 
rather small systems typically composed of 
$N=10^2-10^3$ particles have been used with the periodic boundary 
condition (PBC).
Such small systems have generally been considered to be large enough 
to avoid finite-size effects in the case of amorphas materials in which 
no long-range order exists.
In fact, static properties such as the radial distribution function 
$g(r)$ or the static structure factor $S(q)$ of glassy materials 
are not seriously affected by the system size as long as 
reasonably large systems ($N\gs10^2$) are used. 
However, this is not always the case for dynamical properties.
For example, it is known that the use of a small system with PBC gives 
an manifest effect in relatively short-time behavior of the 
density-time correlation function.
There appears an artifact in time scale of order $L/c$, 
where $L$ is the size of the simulation cell and $c$ is the
sound velocity \cite{Lewis,Hansen}.
As we already mentioned, the dynamical correlation length $\xi$ in 
supercooled liquids grows rapidly with lowering the temperature.
It is thus possible that some kinds of finite-size effects may appear 
in the dynamics of supercooled liquids when $\xi$ becomes comparable 
to $L$ even if no such effect is detected in the 
static correlation functions.
The main purpose of this paper is to examine carefully this point for a
simple soft sphere mixture.
%\\
%\section{Simulations}
%\indent 

Our model mixture is composed of two soft sphere
components $1$ and $2$ having the size ratio $\sigma_1/\sigma_2=1/1.2$
and the mass ratio $m_1/m_2=1/2$ while $\epsilon_1=\epsilon_2=\epsilon$.
The units of length, time, and temperature are $\sigma_1$,
$\tau_0=({m_{1}\sigma_{1}^{2}/\epsilon})^{1/2}$, and $\epsilon/k_B$
in this paper.
Details of simulations are given in our 
earlier paper \cite{Yamamoto3}.
We presently performed MD simulations only in three-dimensional space
with the systems composed of $N=N_{1}+N_{2}=108$, $10^3$, 
and $10^4$ particles while the density $\rho=N/L^3=0.8$ and 
the composition $N_1/N=0.5$ are fixed. 
The corresponding system linear dimensions are 
$L^{N=108}=5.13$, $L^{N=10^3}=10.8$, and $L^{N=10^4}=23.2$.
Simulations were carried out at $T= 0.772$, $0.473$, $0.352$, 
$0.306$ and $0.267$ with the time step $\Delta t = 0.005$.
The PBC was used in all cases.
At each temperature, the systems were carefully equilibrated in the 
canonical condition so that no appreciable aging effect takes place.
Data are then taken in the microcanonical condition.

We first calculate the partial static structure factor,
\be
S_{ab}(q)= \frac{1}{N}\left\langle\sum^{N_a}_{j=1}\sum^{N_b}_{k=1}
\exp({i{\bi q}\cdot{({\bi r}^a_j-{\bi r}^b_k)}})\right\rangle,
\en
to investigate whether finite-size effects are detectable in 
static particle configurations.
Here ${\bi r}^a_{j}$ and ${\bi r}^b_{k}$ 
are the positions of the $j$-th and $k$-th particles in the 
$a$ and $b$ components ($a,b\in1,2$) and $\langle \cdots \rangle$ 
indicates the ensemble average over different initial configurations.
The dimensionless wave number $q$ is in units of $\sigma_1^{-1}$.
In Fig.~1, we plotted $S_{11}(q)$ for $N=108$, $10^3$, and $10^4$ 
at $T=0.473$ (a) and $0.267$ (b).
One can find that $S_{11}(q)$ for all cases excellently 
agrees with each other both in (a) and (b);
no systematic size dependence can be detected among them.
We examined also $S_{12}$ and $S_{22}$ and confirmed 
the same tendency. 
Our results indicate that finite-size effects are very small or 
almost negligible in static pair correlations for $N\gs10^2$,
as that is generally believed.
% At first glance, overall shapes in $S_{11}(q)$ for $T=0.473$ and
% $0.267$ are rather similar to each other. 
% One can, however, find some quantitative differences between them.
% The heights of the first peaks increase at lower temperatures 
% as is usually observed.
% The same tendency has been confirmed for other temperatures.

Let us next consider finite-size effects in dynamical properties.
The structure relaxation in glassy materials can be measured
by calculating the coherent or the incoherent intermediate scattering 
functions, $F(q,t)$ or $F_s(q,t)$.
The decay profiles of those two functions tend to coincide at the first 
peak wave number $q_m$ in $F(q,0)$.
This has been confirmed for the present soft-sphere mixture \cite{Yamamoto3}
and also for a Lennard-Jones binary mixture \cite{Kob2}.
Since $F_s(q,t)$ can be more accurately determined via MD simulation, 
we here calculate the incoherent scattering function for the component 1,
\begin{equation}
F_s(q,t)=\frac{1}{N_1}\Biggl\langle\sum_{j=1}^{N_1} \exp[i{\bi q} \cdot
\Delta{\bi r}^1_j(t)]\Biggr\rangle,
\end{equation}
where
$\Delta{\bi r}^1_j(t)={\bi r}^1_j(t)-{\bi r}^1_j(0)$ is the displacement
vector.
Although $F_s(q,t)$ decays monotonically in normal liquid states, 
it exhibits multi-step relaxations in highly supercooled states.
This is due to the fact that at lower temperatures the particle motions 
are highly jammed and thus trapped considerably in effective cages 
formed by their neighbors.
We then defined the $\alpha$ relaxation time $\tau_{\alpha}$, 
which corresponds to a characteristic life time of the effective cage, 
by $F_s(q,\tau_{\alpha})=e^{-1}$ at $q=2\pi$ for several temperatures.
Fig.~2 shows the decay profiles of $F_s(q=2\pi,t)$ obtained for 
$N=108$, $10^3$, and $10^4$ at $T=0.473$ and $0.267$.
At $T=0.473$, we see that the two curves from $N=10^3$ and $10^4$ 
entirely coincide, and one from $N=108$ is also close to them.
The relaxation times thus obtained are $\tau_{\alpha}^{N=108}\simeq 3.5$, 
$\tau_{\alpha}^{N=10^3}=\tau_{\alpha}^{N=10^4}\simeq 2.3$.
However, the situation is different at $T=0.267$ where $F_s(q,t)$ 
exhibits two-step relaxations.
The faster and the slower parts of the decay are called 
the fast-$\beta$ (thermal) and the $\alpha$ relaxations, respectively.
We note that the decay profiles for the three
systems differ significantly in the $\alpha$ regime ($t\gs10^2$),
whereas they agree well in the fast-$\beta$ regime ($t\ls1$).
Here we determined $\tau_{\alpha}^{N=108} \simeq 11000$,
$\tau_{\alpha}^{N=10^3} \simeq 6500$, and $\tau_{\alpha}^{N=10^4} 
\simeq 2000$ at $T=0.267$.
Fig.~3 shows the temperature dependence of $\tau_{\alpha}^{N=108}$, 
$\tau_{\alpha}^{N=10^3}$ and $\tau_{\alpha}^{N=10^4}$.
At the highest temperature $T=0.772$, $\tau_{\alpha}^{N=108}$,
$\tau_{\alpha}^{N=10^3}$ and $\tau_{\alpha}^{N=10^4}$ are exactly equal.
However, $\tau_{\alpha}^{N=108}$ begins to deviate from the others around
$T=0.473$ at which $\xi\simeq 5$ \cite{Yamamoto1} is comparable to 
$L^{N=108}=5.13$.
The deviation increases with further 
decreasing temperature, and $\tau_{\alpha}^{N=108}$ becomes 
almost one order larger than $\tau_{\alpha}^{N=10^4}$ 
for $T\ls0.352$.
Furthermore, $\tau_{\alpha}^{N=10^3}$ begins to deviate from
$\tau_{\alpha}^{N=10^4}$ around $T=0.306$, at which 
$L^{N=10^3}<\xi<L^{N=10^4}$.
We thus suppose that the present finite-size effects are attribute to 
suppressions of cooperative particle motions due to insufficient 
system size.
The structural relaxation time of smaller systems thus tend to show a 
stronger (super-Arrhenius) temperature dependence as a result of 
the finite-size effects.
Remembering the fact that the static structure factors are almost 
identical among those three systems at all temperatures, the origin of 
this effect may be purely kinetic, or higher order correlations 
in particle configurations may relevant to this.
In their recent paper \cite{Horbach}, Horbach {\it et al.} found similar 
finite-size effects in a model silica glass which is known as a 
typical {\it strong} glass former, in addition to the present soft sphere
mixture which is usually classified in {\it fragile} glass former.

To understand what happens in microscopic scale, we next visualize
individual particle motions in $N=10^4$ system at $T=0.267$.
First, we pick up mobile particles 
by the condition $|\Delta{\bi r}^a_j(t)|>l^a_c$
in a time interval $[t_0,t_0+t]$, where
$t=0.125\tau_\alpha=250$, and $l^a_c$ is defined 
separately for the component $a\in1,2$ such that the sum of 
$\Delta{\bi r}^a_j(t)^2$ over the mobile particles 
covers $66\%$ of the total sum $\sum_{j}^{N_a}\Delta{\bi r}^a_j(t)^2$.
Then we define clusters of the mobile particles by connecting 
$i$ and $j$ if
$|{\bi r}_i(t)-{\bi r}_j(0)|<0.3(\sigma_{i}+\sigma_{j})$ or
$|{\bi r}_i(0)-{\bi r}_j(t)|<0.3(\sigma_{i}+\sigma_{j})$
similar to Donati {\it et al.} \cite{Donati}.
In Fig.~4, we show spatial distribution of the clusters having the
size $n\ge 5$; those are all chain-like \cite{Donati} and have 
large scale correlations.
Although only $5$\% of the total particles are shown in Fig.~4, 
the sum of $\Delta{\bi r}_j(t)^2$ covers approximately $40$\%
of the total $\sum_{i=1}^{N}\Delta{\bi r}_j(t)^2$.
This clearly indicates that the cooperative motions become 
dominant in glassy states.
To investigate finite-size effects in cooperative motions 
quantitatively, we here introduce the distribution function,
\be
P(n)=\sum_{i=1}^{N}\left.^{\prime}\right.\Delta{\bi r}_j(t)^2 \delta(n-n_i)
\left/ \sum_{i=1}^{N}\left.^{\prime}\right.\Delta{\bi r}_j(t)^2 \right.,
\en
where 
the sum runs over mobile particles only.
$n_i$ is the size of the cluster in which 
the mobile particle $i$ belongs, 
and thus $\delta(n-n_i)$ is $1$ if $i$ 
is a member of the cluster having the size $n$ and $0$ if not.
The physical meaning of $P(n)$ is as follows; clusters having the size 
$n$ contribute $P(n)$ to the total squared displacements of the mobile 
particles.
In Fig.~5, we show $P(n)$ for $N=108$, $10^3$, and $10^4$ at $T=0.267$
at which $\xi\simeq40$ obtained for $N=10^4$ is even larger than 
$L^{N=10^4}=23.2$.
We found that the cooperative motions in $N=108$ system are strongly 
suppressed.
By comparing $N=10^3$ and $10^4$, it is found that larger scale 
cooperative motions ($n>10$) are considerably suppressed also in 
$N=10^3$ system.
Infinitely large clusters which percolate the system though the PBC
have never been found in all cases.
The characteristic cluster size $\bar{n}=\sum_{i=1}^\infty nP(n)$ thus 
obtained is 3.38, 6.11, and 7.73 for $N=108$, $10^3$, and $10^4$, 
respectively.

In the framework of conventional liquid theories \cite{Hansen}, 
changes in static particle configurations upon cooling
lead to a drastic slowing-down toward the grass transition.
The mode coupling theory (MCT) \cite{mode1,mode2} is the most successful 
self-consistent approach within this framework.
MCT describes onset of glassy slowing down (or slow structural 
relaxations) in the density-time correlation functions.
In the original MCT, however, a sharp Ergodic/Non-ergodic transition 
is predicted at a temperature $T_0$ which is considerably above $T_g$.
Although such a tendency has been found in colloidal systems in which
thermal activation processes are negligible, it has never been observed 
in real glassy materials.
It is thus believed that the MCT has some difficulties for describing 
the true dynamics of supercooled liquids apparently below $T_0$.
The main problem is the fact that the original MCT do not take into 
account the hopping motions of particles, which must be cooperative 
and thus long ranged as is seen in recent MD simulations
\cite{Yamamoto1,Yamamoto3,Muranaka,Perera,Kob,Donati}.
Unfortunately the problem has not yet been overcome
in fully self-consistent way because efforts for including thermal 
activations make the theory more or less {\it ad hoc}.
It is a interesting fact that the behavior of structural relaxations
in our smallest system, in which cooperative hopping motions are highly 
suppressed, becomes somehow closer to the original MCT prediction.

It is worth mentioning several experimental attempts to 
find the evidence of the dynamical heterogeneity in glassy materials.
One of the most interesting and useful approaches 
are the recent experiments on glass-forming thin films.
The thickness $d$ dependence of film properties is the main interest 
in these studies \cite{Wallace,Forrest,Jerome,Fukao}.
%J\'er\^ome and Commandeur \cite{Jerome} investigated the relaxation behavior
%of a glass-forming liquid crystal confined between two walls.
%Fukao and Miyamoto \cite{Fukao} measured the thickness dependence of the
%glass transition temperature $T_g$ of polymeric films.
The motivations of those studies are quite similar to the present study; 
they aimed to control the size $\xi$ of the cooperative motions by changing 
$d$ and found that the relaxation time and $T_g$ considerably depend
on $d$.
They considered that the cooperative particle motions in the direction 
normal to the film may be truncated near the interface, and this effect
become dominant when $\xi\gs d$.
Thus the system size restriction can {\it enhance} the particle motions. 
Note that this seems to contradict to the finite-size effect 
in the present MD 
simulations in which the size restriction {\it suppresses} 
the cooperative particle motions \cite{FSEMD}.
The mechanism of this discrepancy is still an open question; we naively 
speculate that the situation is much more complicated in polymer films 
than in MD simulations.
The system size restriction occurs only in one direction normal to the 
film and other two in-film directions are free in thin films,
whereas all directions are equally restricted in the present simulations.
We believe that investigating the microscopic relaxation mechanisms in 
glass-forming (both simple and polymeric) thin films are definitely 
important.

In summary, we have examined system-size effects in the dynamics of 
supercooled liquids by MD simulation.
We found significant finite-size effects in the structural relaxation 
at lower temperatures, whereas no such effect is detectable in static pair 
correlation functions.
The cooperative particle motions, which leads to the $\alpha$ 
relaxation in glassy states, are strongly suppressed in 
smaller systems for temperatures lower than $T_c$ at which $\xi$ 
becomes comparable to the system size.
The present finite-size effects are regarded as a natural consequence 
of the dynamical heterogeneity appearing in supercooled liquids.
We point out that finite-size effects are significant in the dynamics 
of highly supercooled liquids, and thus special attention must be 
payed to investigate true relaxation dynamics in computer simulations 
particularly in the $\alpha$ regime.

We thank Profs. T. Munakata and A. Onuki for helpful discussions.
This work is supported by a Grant in Aid for Scientific 
Research from the Ministry of Education, Science, Sports and Culture
of Japan.
Calculations have been carried out at the Supercomputer 
Laboratory, Institute for Chemical Research, Kyoto University 
and Human Genome Center, Institute of Medical Science, 
University of Tokyo.

%%%%% Fig.1 %%%%%%%%%%%%%%%%%%
\begin{figure}[t]
\epsfxsize=3.in
\centerline{\epsfbox{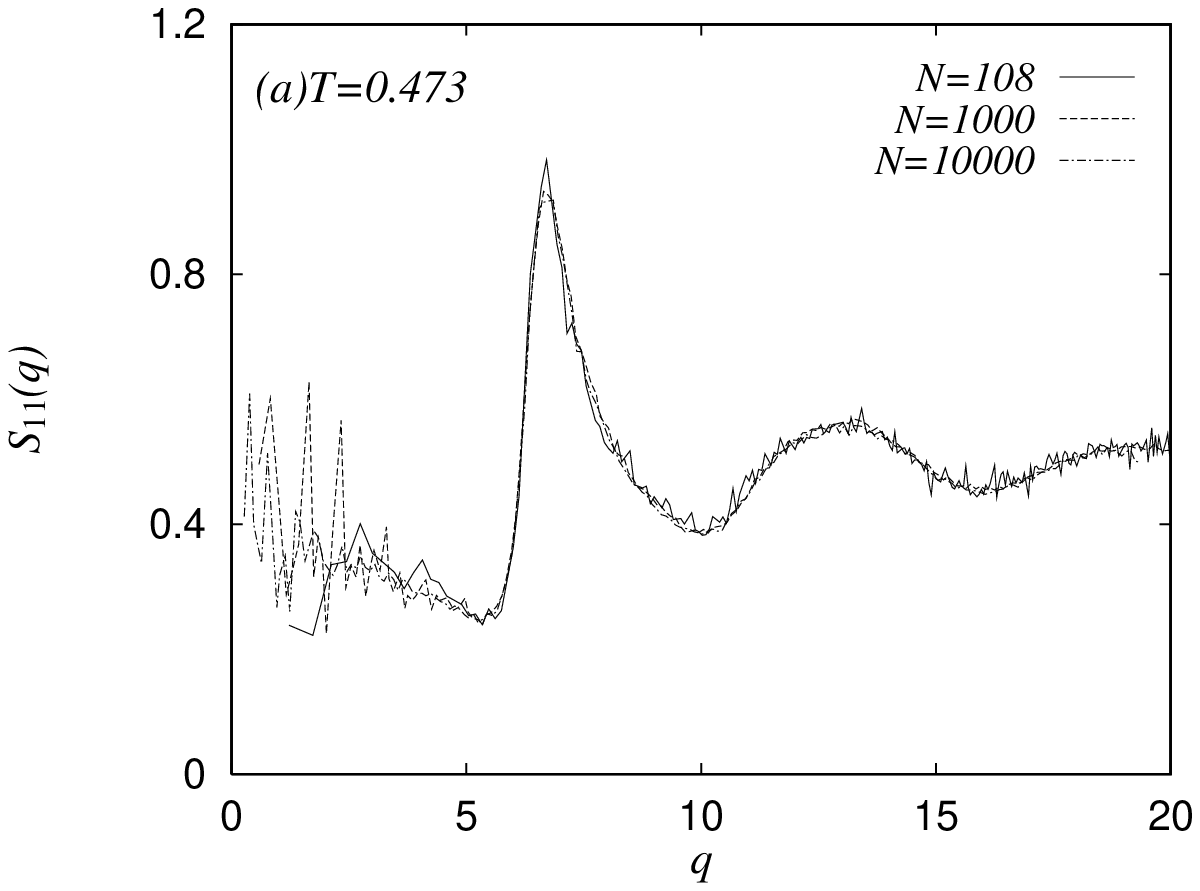}}
\epsfxsize=3.in
\centerline{\epsfbox{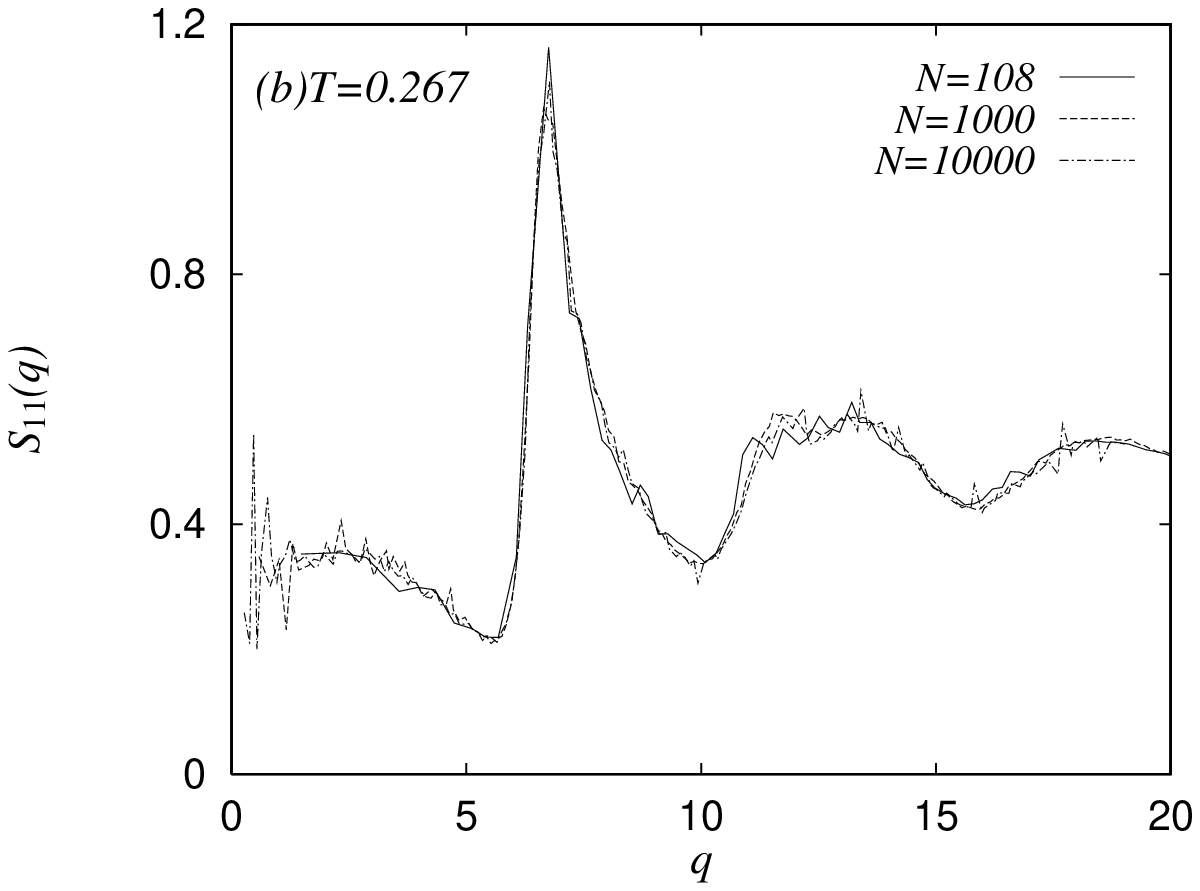}}
\caption{\protect\narrowtext
The partial static structure factor $S_{11}(q)$ 
obtained at $T=0.473$ (a) and $0.267$ (b)
for $N=108$, $10^3$, and $10^4$ systems.
}
\label{fig1}
\end{figure}
\noindent
\vspace{-5mm}

%%%%% Fig.2 %%%%%%%%%%%%%%%%%%
\begin{figure}[t]
\epsfxsize=3.in
\centerline{\epsfbox{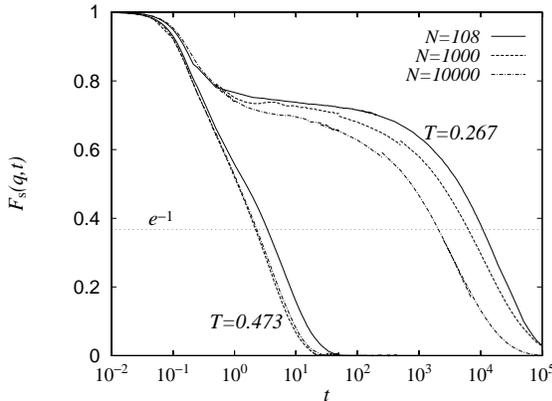}}
\caption{\protect\narrowtext
The incoherent intermediate scattering function $F_s(q,t)$ 
of the component $1$ with $q=2\pi$ at $T=0.473$ and $0.267$.
}
\label{fig2}
\end{figure}
\noindent
\vspace{-5mm}

%%%%% Fig.3 %%%%%%%%%%%%%%%%%%
\begin{figure}[t]
\epsfxsize=3.in
\centerline{\epsfbox{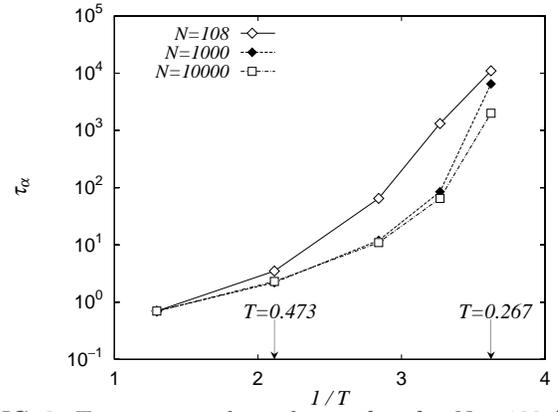}}
\caption{\protect\narrowtext
Temperature dependence of $\tau_{\alpha}$ for $N=108$ (open diamonds), 
$10^3$ (filled diamonds), and $10^4$ (open squares).
}
\label{fig3}
\end{figure}
\noindent
\vspace{-5mm}

%%%%% Fig.4 %%%%%%%%%%%%%%%%%%
\begin{figure}[t]
\epsfxsize=3.in
\centerline{\epsfbox{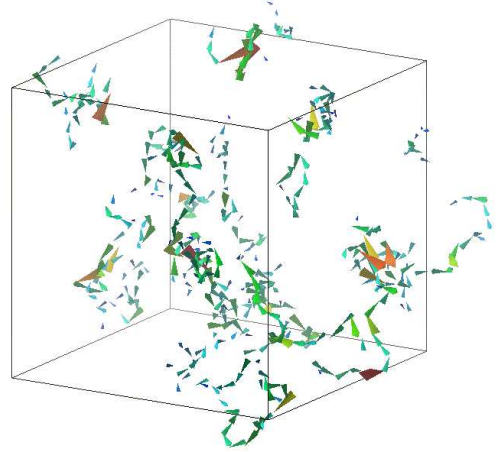}}
\caption{\protect\narrowtext
Spatial distribution of particle displacements having the 
cluster size $n\ge 5$ at $T=0.267$ for $N=10^4$.
The arrows indicates individual particle displacements.}
\label{fig4}
\end{figure}
\noindent
\vspace{-5mm}

%%%%% Fig.5 %%%%%%%%%%%%%%%%%%
\begin{figure}[t]
\epsfxsize=3.in
\centerline{\epsfbox{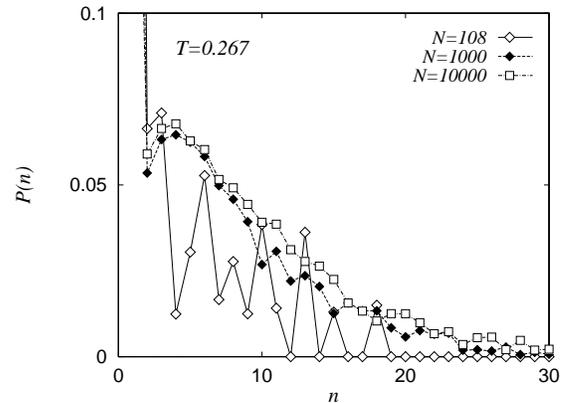}}
\caption{\protect\narrowtext
$P(n)$ versus $n$ at $T=0.267$ for $N=108$ (open diamonds), 
$10^3$ (filled diamonds), and $10^4$ (open squares).
}
\label{fig5}
\end{figure}
\noindent
\end{multicols}


\vspace{-5mm}
\begin{references}
\vspace*{-10mm}

%Email address: kin@kuamp.kyoto-u.ac.jp
%Email address: ryoichi@ton.scphys.kyoto-u.ac.jp

%Review
\bibitem{Jackle}
J. J\"ackle, Rep. Prog. Phys. {\bf 49}, 171 (1986).
%\bibitem{Ediger}
%M. D. Ediger, C. A. Angell, and R. Nagel, J. Phys. Chem. {\bf 100}, 
%13200 (1996).
\bibitem{Tohwa}
{\it Slow Dynamics in Complex Systems}, 
edited by M. Tokuyama and I. Oppenheim,
AIP Conference series 469, (AIP, New York, 1999).

%Early MD
\bibitem{Bernu} B. Bernu, Y. Hiwatari and J.P. Hansen,
J. Phys. C {\bf 18}, L371 (1985); 
B. Bernu, J.P. Hansen, Y. Hiwatari and G. Pastore 
Phys. Rev. A  {\bf 36}, 4891 (1987); 
J. Matsui, T. Odagaki and Y. Hiwatari, 
Phys. Rev. Lett. {\bf 73}, 2452 (1994). 

%Our MD
\bibitem{Yamamoto1}
R. Yamamoto and A. Onuki, J. Phys. Soc. Jpn. {\bf 66}, 2545 (1997);
Europhys. Lett. {\bf 40}, 61 (1997);
A. Onuki and R. Yamamoto, J. Non-Cryst. Solids, {\bf 235 -237}, 34-40 (1998);
R. Yamamoto and A.Onuki, Phys. Rev. E {\bf 58}, 3515 (1998).
\bibitem{Yamamoto3}
R. Yamamoto and A.Onuki, 
Phys. Rev. Lett., {\bf 81}, 4915-4918 (1998);
in Ref. \cite{Tohwa}, 476-483.

%Other recent MD
\bibitem{Muranaka}
T. Muranaka and Y. Hiwatari, Phys. Rev. E {\bf 51}, R2735 (1995);
J. Phys. Soc. Jpn. {\bf 67}, 1982 (1998)
\bibitem{Perera}
D.N. Perera and P. Harrowell, Phys. Rev. E {\bf 54}, 1652 (1996);
D.N. Perera, J. Phys.: Condens. Matter, {\bf 10}, 10115-10134 (1998).
\bibitem{Kob}
W. Kob {\it et al.}, Phys. Rev. Lett. {\bf 79}, 2827 (1997);
S. Glotzer and C. Donati, 
J. Phys.: Condens. Matter, {\bf 11}, A285-A295 (1999).
\bibitem{Donati}
C. Donati {\it et al.},
Phys. Rev. Lett. {\bf 80}, 2338 (1998).

\bibitem{Lewis}
L.J. Lewis, G. Wahnstr\"om, Phys. Rev. E {\bf 50}, 3865 (1994).
\bibitem{Hansen}
J.P. Hansen and I.R. McDonald, {\it Theory of Simple Liquids} 
(Academic Press, London, 1986).

\bibitem{Kob2}  W. Kob and H.C. Andersen, 
Phys. Rev. E {\bf 52}, 4134 (1995).

%System size effect
\bibitem{Horbach}
J. Horbach, W. Kob, K. Binder and C.A. Angell, Phys. Rev. E {\bf 54},
 R5897 (1996).
%J. Horbach, W. Kob and  K. Binder, cond-mat/9901162.


%Liquid theory
\bibitem{mode1}
U. Bengtzelius, W. G$\ddot{\rm o}$tze and A. Sj$\ddot{\rm o}$lander,
J. Phys. C {\bf 17},  5915 (1984).
%W. G\"otze and L. Sj\"ogren, Rep. Prog. Phys. {\bf 55}, 241 (1992).
\bibitem{mode2} E. Leutheusser,
Phys. Rev. A {\bf 29}, 2765 (1984).


%Thin films
\bibitem{Wallace}
W.E. Wallace, J.H. Van Zanten and W.L. Wu, Phys. Rev E {\bf 52}, 
R3329 (1995).
\bibitem{Forrest}
J.A. Forrest, K. Dalnoki-Veress and J.R. Dutcher, Phys. Rev. E {\bf 56}, 
5705 (1997).
\bibitem{Jerome}
B. J\'er\^ome and J. Commandeur, Narure {\bf 386}, 589 (1997);
B. J\'er\^ome, J. Phys.: Condens. Matter, {\bf 11}, A189-A197 (1999).
\bibitem{Fukao}
K. Fukao and Y. Miyamoto, Europhys. Lett. {\bf 46}, 649 (1999).

\bibitem{FSEMD} Having slower relaxations in smaller systems is a 
general trend in MD simulation as seen also in 
S. B\"uchner and A. Heuer, cond-mat/9906280 and \cite{Horbach}.

\end{references}
\end{document}